# Fundamental thermal noise in droplet microresonators

A. Giorgini,[1] S. Avino,[1] P. Malara,[1] P. De Natale,[2] G. Gagliardi[1]*

[1]*Consiglio Nazionale delle Ricerche, Istituto Nazionale di Ottica (INO), via Campi Flegrei 34, Complesso "A. Olivetti", 80078 Pozzuoli (Napoli), Italy*
[2]*Consiglio Nazionale delle Ricerche, Istituto Nazionale di Ottica (INO), Largo E. Fermi 6, 50125 Firenze, Italy*

Liquid droplet whispering-gallery-mode microresonators open a new research frontier for optomechanics and photonic devices. At visible wavelengths, where most liquids are transparent, a major contribution to a droplet optical quality factor is expected theoretically from thermal surface distortions and capillary waves. Here, we investigate experimentally these predictions using transient cavity ring-down spectroscopy. In this way, the optical out-coupling and intrinsic loss are measured independently while any perturbation induced by thermal, acoustic and laser-frequency noise is avoided thanks to the ultra-short light-cavity interaction time. The measurements reveal a photon lifetime ten times longer than the thermal limit and suggest that capillary fluctuations activate surface scattering effects responsible for light coupling.

**Keywords**: whispering-gallery mode, liquid droplet, cavity ring-down, photon lifetime, Q-factor, capillary waves

Dielectric micro-resonators obtained from solid glasses with different geometries have proven very promising devices for bio-chemical sensing [1], optical-frequency comb generation [2] and cavity opto-mechanics [3]. The sample delivery modalities and the weak interaction with the whispering-gallery-mode (WGM) evanescent tail severely limit their feasibility as sensors in liquid environments and integration in microfluidic systems. The possibility to form liquids into small, easy-to-handle droplet microresonators has triggered a strong scientific interest towards the creation of novel versatile photonic elements, i.e. a lab-in-a-droplet platform [4-7]. The long-standing issue of direct sensing in liquid media may be overcome in an elegant and effective way as a liquid resonator comprises the sensing unit and the sample under investigation at the same time. On the other hand, droplets are extremely appealing as opto-mechanical oscillators as they are among the simplest optofluidic devices and are made of materials $10^6$ times softer than glasses, which increase the amplitude of Brownian fluctuations and capillary waves [8,9]. Excitation of high-quality WGMs in droplets have been recently demonstrated and advanced techniques from the realm of standard cavities have been extended to liquids for lasing, spectroscopy, sensing and manipulation [10-14]. Analogously to solid microcavities, the quality factor is the primary figure of merit, which carries information on the photon lifetime within the system. The main limitations usually arise from material loss, scattering and coupling mechanism.

In this regard, besides absorption and scattering, also liquid surface tension is crucial to the optical features of droplets. Indeed, thermally-induced shape distortions restored by surface tension give rise to capillary waves on the surface that typically occur at frequencies in the 0.1-1 MHz range, depending on the size and material [15,16]. They manifest as a fast jitter of the optical resonances [16,17] and thus inevitably affect the Q-factor when observed by a direct spectroscopic measurement. Surface light scattering is strongly influenced by capillary waves which can be used as a direct way to measure surface tension and viscosity of fluids [18,19]. In addition, similar effects may be caused by optical-to-mechanical energy transfer or by light impulses that induce acoustic waves via electrostriction [20-21]. It is still not obvious to what extent this kind of opto-mechanical feedback influences the Q-factor of microresonators and whether it is detrimental or not [16]. All these physical processes affecting the quality factor are not only crucial for characterization of droplet microresonators as liquid sensors but also for their use as miniaturized laboratories to study surface tension and rheological properties of materials via light scattering on the liquid-air interface [19]. In our work, we investigate experimentally the Q-factor of various oil-droplet microresonators with regard to the limits dictated by scattering, thermal noise and optical absorption loss, using a free-space excitation scheme to fully harness the benefits of a liquid cavity. We perform cavity ring-down measurements in its relaxation regime that provide information on the photon lifetime, and determine the effective Q-factor influenced by the coupling mechanism, i.e. the loaded Q-factor. We find that light scattering due to surface thermal capillary waves serves as an optical access to the whispering-gallery modes. We demonstrate that the Q-factor of highly-transparent, vertically-suspended oil spheres can be higher than $10^7$, in the visible region, revealing the intrinsic cavity photon lifetime that in most experiments is masked by thermal effects. Moreover, our results show that the ultimate Q-factor is not limited by thermal surface fluctuations as opposed to theoretical predictions.

As shown in Fig. 1, a DFB diode laser emitting at a wavelength of 640 nm is used to interrogate the liquid cavities. The laser is driven by a precision temperature controller and a low-noise current generator. A paraffin oil droplet, suspended by the tip of an acrylate-coated silica fiber (250-µm outer diameter) owing to surface tension [6], serves as the optical cavity. Its radius $a$ can range approximately from 200 µm to 600 µm depending on the minimum volume required for a stable, symmetrical droplet and the maximum weight allowed by the equilibrium between surface tension and gravity force. A fine alignment of the droplet into the laser beam is done by a piezo-actuated xyz translation stage with 20-nm resolution while a microscope objective focuses the light beam tangential to the sphere rim in order to excite WGMs. In our configuration, the beam focus optimal position is just inside the surface edge and the cavity is undercoupled with 5-10 % of the incident power effectively injected into low-order WGMs. Basically, the main optical coupling mechanisms are believed to be surface scattering [22,23] and chaos-assisted tunneling [24], whereas evanescent-wave coupling to WGM in free-space is not relevant due to the small droplet curvature [25,26]. Moreover, as remarked in [27], van de Hulst's localization principle fails in the case of a droplet with surface irregularities. Small irregularities caused by thermal capillary waves induce angular-momentum coupling of partially-scattered light waves to the resonator WGMs with optimal excitation efficiency for a light beam focused slightly inside the surface [27,28], in agreement with experimental findings. Direct transmission and scattering WGMs are collected by additional lenses and focused on separate photodiodes. A current scan is sent to the laser to observe several WGM spectra. Fig. 2 shows three of the narrowest resonances detected over a cavity free-spectral-range, where positive peaks correspond to the scattered light while negative dips appear in the transmitted light. The signal spectra are highly repeatable and the alignment of the droplet is not critical to the quality factor as opposed to tapered-fiber or prism coupling schemes [29]. Thanks to free-space beam illumination, a number of resonances are simultaneously excited in the liquid cavity, being likely associated to WGMs with low and high radial orders. In addition, the elliptical geometry of the suspended droplet potentially splits WGMs among different azimuthal orders [3]. A comprehensive theoretical description of the scattering efficiency and internal intensities for any dielectric sphere can be found in [30], reproducing quite well the spectra that we experimentally observed. Droplet cavities made of other oils or water can be obtained with similar Q-factors but paraffin and silicone oil have been preferred for this study owing to their low evaporation rate and high transparency.

The Q-factor of the WGM of a droplet, with optical frequency ν, can be expressed in terms of the different loss channels that contribute to reduce the overall photon lifetime τ as [31]

$$\frac{1}{Q} = \frac{1}{Q_{\text{abs}}} + \frac{1}{Q_{\text{rad}}} + \frac{1}{Q_{\text{ss}}} + \frac{1}{Q_{\text{shape}}} \tag{1}$$

where $Q = 2\pi\nu\tau$ and $Q_{abs}$, $Q_{rad}$, $Q_{ss}$, $Q_{shape}$ are the contributions of material absorption, radiative loss, surface scattering loss and thermal shape distortions, respectively. A relative decay rate can be associated to each of the above terms. As remarked above, in microresonators with radii of the order of few hundreds µm, radiative loss is negligible compared to surface scattering loss, which in fact is the main in-out coupling mechanism. However, the factors related to absorption and thermal noise dominate the quality budget of our liquid microresonator. In particular, shape distortions due to thermally-induced capillary waves on the liquid surface have clearly been evidenced by current theoretical models as an inherent limitation to the observed Q-factor [16]. Indeed, they modulate the splitting of quasi-degenerate azimuthal modes due to a dynamical change in the droplet eccentricity thus leading to an effective WGM line broadening. The corresponding reduction of quality factor, considering the dependence of the azimuthal mode frequency on the droplet eccentricity and assuming a WGM with the lowest radial order and maximum angular momentum $l \approx 2\pi n a/\lambda$ [17], is

$$Q_{shape} \cong \frac{2a}{\Delta} \qquad (2)$$

where $\Delta$ is the root-mean-square amplitude of the thermally-induced droplet fluctuations. The latter is given by a recent theoretical study about surface displacements on a spherical liquid-air interface [32]

$$\Delta = \sqrt{\frac{k_B T}{4\pi\gamma} \ln\left(\frac{2a^2}{3\sigma^3}\right)} \qquad (3)$$

where $k_B$, $\gamma$, and $T$ are the Boltzmann constant, the liquid surface tension and the absolute temperature, respectively, while $\pi\sigma^2$ is the effective area per molecule [32]. If we consider $a = 450$ µm as an average value within our size range, $T = 296$ K, $\gamma = 0.026$ N/m, and $\sigma \approx 6\cdot10^{-10}$ m (calculated from the average molecular radius with a molecular weight ∼ 500 for oils), the limit set by thermal fluctuations in Eq. (2) would be $Q^*_{shape} = 1.6\cdot10^6$ for paraffin and other mineral oils.

The material absorption contribution to the cavity Q-factor can instead be expressed as

$$Q_{abs} = \frac{2\pi n}{\alpha\lambda} \qquad (4)$$

where $n$ and $\alpha$ are the liquid refractive index and absorption coefficient at the laser wavelength $\lambda$. For paraffin and mineral oil ($n = 1.47$, $\alpha \approx 0.1$ m$^{-1}$ [33,34]) $Q_{abs}$ results to be ∼ $1.4\cdot10^7$ while for silicone oil this factor is expected to be higher due to its very small optical absorption [10]. Heavy water (D$_2$O) would exhibit the lowest absorption coefficient around 640 nm with $\alpha \approx 0.02$ m$^{-1}$ [35].

Experimentally, the Q-factor of an optical cavity can be determined in different ways, the easiest of which relies on a spectroscopic measurement of the WGM optical frequency-to-linewidth ratio. However, this kind of measurement is highly inaccurate as it requires a non-linear fit of the WGM resonance profile that, in most cases, is seriously affected by laser amplitude noise, laser-cavity frequency jitter and photothermal broadening while it depends significantly on the observation time [25,36,37]. A time-domain measurement of photon lifetime provides instead a direct, reliable, low-noise estimate of the overall cavity loss. This can be performed by cavity ring-down spectroscopy (CRDS), i.e. a measurement of the characteristic time of the exponential decay resulting from fast shutting a resonant laser beam [6,37]. However, a microresonator typically exhibits lifetimes in the order of 1-10 ns that makes ring-down measurements challenging even for high bandwidth modulation devices and detection electronics. Also, it only provides the loaded Q-factor value. An alternative method consists in the so-called rapidly-swept CRDS, where the laser is not constantly resonant whereas it is scanned through the resonance faster than the cavity build-up [37,38]. Because of the time-beating between photons entering the cavity and those previously stored into it, a chirped ringing response emerges on the transmitted beam that carries information on the scan rate, the intrinsic cavity lifetime and the coupling loss rate [39]. The analytical function of the microcavity transmission in a non-stationary condition is given by $T = \left|\frac{E_{out}}{E_{in}}\right|^2$ with $E_{out}$ and $E_{in}$ the output and input cavity fields, respectively, which are related each other by [36]

$$E_{out} = -E_{in} + \sqrt{\frac{2}{\tau_c}} v(t) \tag{5a}$$

and

$$v(t) = \sqrt{\frac{2}{\tau_c}} E_{in} \exp(j\omega_0 - t/\tau) \left[ f(t) - f(0) + \frac{\tau}{1+j(\omega_i-\omega_0)\tau} \right] \tag{5b}$$

$$f(t) = -\sqrt{\frac{j\pi}{2V_s}} \exp\left[-\frac{j(\omega_i-\omega_0-j/\tau)^2}{2V_s}\right] erf\left(\frac{\frac{j}{\tau}+\omega_i-\omega_0-V_s t}{\sqrt{2jV_s}}\right) \tag{5c}$$

where erf(z) with z∈C denotes the complex error function. Here $1/\tau = 1/\tau_c + 1/\tau_0$ is the loaded photon lifetime, where $\tau_c$ is the loss rate associated with the out-coupling and $\tau_0$ is the intrinsic photon lifetime.

In Fig. 3, we show the ringing profiles corresponding to the sharpest resonance on the blue side of Fig. 2 (WGM1) and a neighboring peak (WGM2). They are recorded modulating the laser injection current with a triangular wave at a frequency of 1 MHz, while a 12-GHz photodiode monitors the transmission. As expected, distinct oscillations appear only on the narrowest peak (WGM1). It is worth noting that the scattered power and the transmitted power exhibit a different symmetry in their ringing profiles [38]. The ringing signal of WGM3 is also shown in Fig. 4.

After normalizing the signals by the laser-scan slope, the experimental curves are averaged over 4000 fast-repeated acquisitions (with 300-ns time gap) and fitted with the expressions of Eqs. (5) retrieving lifetimes $\tau_1$ = 6.35±0.05 ns ($Q_1$ = (1.87±0.01) $10^7$) for the narrowest mode and $\tau_3$ = 4.8±0.1 ns ($Q_3$ = (1.41±0.03) $10^7$) for the central mode. In Fig. 4, we show the agreement between the experimental data and a multiple-resonance fit curve that accounts also for the presence of a weak mode nearby. The resulting values are about one order of magnitude larger than the calculated thermal-noise bound $Q^*_{shape}$. We have repeated the same measurements varying the beam-droplet distance within a reasonable resonance-visibility range (± 2.5 μm) without finding any significant change in the lifetimes.

The large measured Q-factors are close to the expectations based on liquid paraffin absorption only. To verify this hypothesis, we have suspended four paraffin-oil droplets with different radii and performed subsequent measurements with the time-ringing technique on the central mode, chosen as the best compromise in terms of signal-to-noise ratio and linewidth. The measured decay times fall between 4.43±0.04 ns and 4.66±0.05 ns, as shown in Fig. 5, leading to Q-factors between (1.30±0.01) $10^7$ and (1.37±0.01) $10^7$. From this measurement set, we note that the lifetime remains constant with the drop radius within 2σ, even when it is doubled. The $\tau$ value changes instead if a different liquid is used, with 5.44±0.05 ns ($Q_s$ = (1.60±0.01) $10^7$) for silicone oil. This argument leads us to conclude that the quality factor is mainly affected by intra-cavity absorption loss. Also it confirms that thermal noise is not the actual limit since the dependence on the droplet radius expressed in Eq. (2) is not found in the experiment. On the contrary, the quality factor $Q = \nu/\Delta\nu$ seen from the spectral width $\Delta\nu$ of WGM resonances on a relatively longer scan (5-ms semi-period), is found to be ~ 3.5·$10^6$. Although low-frequency instabilities affecting the resonance profile prevent a reliable linewidth investigation vs droplet size, this value is close to the limit predicted by Eq. (2) and Eq. (3). The discrepancy with the time-domain measurements can be understood considering that thermally-induced shape distortions occur on a fast timescale [16,17] ($10^{-6}$-$10^{-5}$ s) thus leading to wider resonance profiles along typical laser scans. Such fluctuations instead do not act on the ns-scale ringing signals which exhibit a longer photon lifetime, as results from fitting with Eqs. 5. Also, all the fits consistently yield a value of $\tau_c \approx 7$ μs, i.e. > 1500 times the intrinsic lifetime $\tau_0$, that confirms the droplet cavity does not suffer from any loading effect in the case of free-space excitation. We note that the decay rate $1/\tau_c$ provides a direct estimate of the out-coupling loss, i.e. the scattering loss. This contribution is likely due to the average amplitude of thermally-induced broad-spectrum capillary waves on the droplet surface. The surface scattering factor can be calculated with the formula

$$Q_{ss} = \frac{3n^2(n^2+2)^2}{(4\pi)^3(n^2-1)^{5/2}} \frac{\lambda^{7/2}(2a)^{1/2}}{\Delta^2 B^2} \qquad (6)$$

proposed and experimentally tested in [41] for solid microresonators in the 600-nm wavelength range, where $n$ is the liquid refractive index and $B$ is the surface-irregularity correlation length. Taking $B \sim 6$ nm [42] and the above found value of $\Delta$, we obtain $Q_{ss} \sim 2 \cdot 10^{10}$ corresponding to a lifetime $Q_{ss}/2\pi\nu \sim 6.9$ μs, in satisfactory agreement with the value of $\tau_c$ retrieved from our ringing measurements.

In principle, other physical mechanisms might be playing a role in the microresonator loss budget. According to the general theory [16], the resonance linewidth may deviate from the thermal-noise limit of Eq. (3) due to imperfect excitation of the spectral multiplet within the WGM or in the presence of non-linear scattering phenomena that may even cause line-narrowing. While we doubt that only a partial overlap of the laser with the WGM occurs in our case, the second scenario is possible in principle. In Fig. 6, we show the very strong thermally-induced nonlinearity observed when scanning the laser across a WGM, which points out a hysteretic behaviour on the ascending/descending sides of the scan [25]. Strong frequency fluctuations are visible on the resonance lineshape. These effects are more and more relevant when going to slower scan speeds and eventually prevents us to take WGM spectra for linewidth measurements longer than few ms at full laser power. In order to find out whether such strong energy transfer to the cavity is accompanied by non-linear variations of the Q-factor, we have performed a new set of ringing cavity ring-down measurements by gradually decreasing the incident power down to one order of magnitude. However, no correlated change was observed in the photon lifetime. Next, we have frequency-locked the laser on one of the narrowest resonances using an all-optical feedback loop [13] and we have calculated the fast-Fourier transform (FFT) of the transmission and scattering signals. From the FFT noise spectra only an overall increase in the amplitude noise is found on resonance without any noticeable spectral feature associated to high-frequency vibrations or surface waves. Therefore, no conclusive evidence of non-linear optical effects in liquid droplets has been found.

Our experimental results provide evidence that, on a time-scale shorter than thermal fluctuations, the Q-factor of free-space coupled liquid microresonators can be as high as their solid counterpart. Despite theoretical models, we experimentally demonstrate that the actual limitation to the photon lifetime is material absorption, which points to a quality factor $> 10^8$ for highly transparent liquids. Indeed, when observing the laser passage through resonance along a timescale shorter than typical cavity distortions, a measurement of the ringing decay discloses the intrinsic cavity lifetime *unaffected* by thermal and mechanical noise. As a consequence, we believe that no Q enhancement is taking place but rather that the true lifetime is observed instead of an effective one. We find in turn that surface scattering due to thermally-induced fluctuations plays a crucial role for light coupling and we confirm that large beam misalignments along with power or size changes do not affect coupling efficiency. This suggests that droplet microresonators can be a convenient and robust optical platform for spectroscopy and bio-sensing in the liquid phase as well as for investigating Brownian motion, viscoelasticity, surface capillarity phenomena and droplet coalescence in a controlled environment [19,43, 44].

* gianluca.gagliardi@ino.it

1. M. R. Foreman, J. D. Swaim, and F. Vollmer, Adv. Opt. Phot. **7**, 168 (2015
2. T. J. Kippenberg, R. Holzwarth, S. A. Diddams, Science 332, 555 (2011)
3. M. Aspelmeyer, T. J. Kippenberg, F. Marquardt, Rev. Mod. Phys. 86, 1391 (2014)
4. M. Hossein-Zadeh, K. J. Vahala, Opt. Express **14**, 10800 (2006)
5. A. Jonáš, M. Aas, Y. Karadag, S. Manioğlu, S. Anand, D. McGloin, H. Bayraktar, and A. Kiraz, Lab Chip **14**, 3093 (2014)
6. S. Avino, A. Krause, R. Zullo, A. Giorgini, P. Malara, P. De Natale, H.-P. Loock, and G. Gagliardi, Adv. Opt. Mat **2**, 1155 (2014)
7. S. Maayani, L. Martin and T. Carmon, Nat. Comm. **7**, 10435 (2016)
8. A. Doerr, M. Tolan, W. Prange, J-P. Schlomka, T. Seydel, W. Press, D. Smilgies, and B. Struth, Phys. Rev. Lett. **83**, 3470 (1999)
9. C. Fradin, A. Braslau, D. Luzet, D. Smilgies, M. Alba, N. Boudet, K. Mecke, ad J. Daillant, Nature 403, 871 (2000)
10. C. Heinisch, J. B. Wills, J. P. Reid, T. Tschudia and C. Tropea, Phys. Chem. Chem. Phys. **11**, 9720 (2009)
11. M. R. Foreman, S. Avino, R. Zullo, H.-P. Loock, F. Vollmer, and G. Gagliardi, Eur. Phys. J. Special Topics **223**, 1971 (2014)
12. S. Anand, M. Eryürek, Y. Karadağ, A. Erten, A. Serpengüzel, A. Jonáš, and A. Kiraz, J. Opt. Soc. Am. B **33**, 1349 (2016)
13. R. Zullo, A.Giorgini, S. Avino, P. Malara, P. De Natale, and G. Gagliardi, Opt. Lett. **41**, 650 (2016)
14. R. Dahan, L. L. Martin, T. Carmon, Optica **3**, 175 (2016)
15. S. Maayani, L. L. Martin, S. Kaminski, and T. Carmon Optica, **3**, 552 (2016)
16. H. M. Lai, P. T. Leung, and K. Young, Phys. Rev. A **41**, 5199 (1990)
17. H. M. Lai, P. T. Leung, K. Young, P. W. Barber, S. C. Hill, Phys. Rev. A 41, 5187 (1990)
18. D. M. Buzza, Langmuir **18**, 7583 (1997)
19. F. Behroozi, J. Smith, and W. Even, Am. J. Phys. **78**, 1165 (2010)
20. H.-M. Tzeng, M. B. Long, R. K. Chang, P. W. Barber, Opt. Lett. **10**, 209 (1985)
21. J-Z. Zhang and R. K. Chang, Opt. Lett. **13**, 916 (1988)
22. H.-B. Lin, J- D. Eversole, A. J. Campillo, J. P. Barton, Opt. Lett. **23**, 1921 (1998)
23. L. Shao, X.-F. Jiang, X.-C. Yu, B.-B. Li, W. R Clements, F. Vollmer, W. Wang, Y.-F. Xiao, Q. Gong, Adv. Mat. **25**, 5616 (2013)
24. Z. Ballard, M. D. Baaske and F. Vollmer, Sensors **15**, 8968 (2015)
25. V.B. Braginsky, M.L. Gorodetsky, V.S. Ilchenko, Phys. Lett. A **137**, 393 (1989)
26. N. Gaber, M. Malak, X. Yuan, K. N. Nguyen, P. Basset, E. Richalot, D. Angelescu and T. Bourouin, Lab Chip **13**, 826 (2013)
27. J. A. Lock, Opt. Lett. **24**, 427 (1999)
28. J. P. Barton, Appl. Opt. **34**, 5542 (1995)
29. J. H. Chow, M. A. Taylor, T. T-Y. Lam, J. Knittel, J. D. Sawtell-Rickson, D. A. Shaddock, M. B. Gray, D. E. McClelland, and W. P. Bowen, Opt. Express. **20**, 12622 (2012)
30. P. R. Conwell, P. W. Barber, C. K. Rushforth, J. Opt. Soc. Am. A **1**, 62 (1984)
31. A. N. Oraevsky, Quant. Electron. **32**, 377 (2002)
32. L. F. Phillips, J. Phys. Chem. B **104**, 2534 (2000)
33. R. L. P. van Veen, H. J. C. M. Sterenborg, A. Pifferi, A. Torricelli, E. Chikoidze, R. Cubeddu, J. Biomed. Opt. **10**, 054004 (2005)
34. A. R. Chraplyvy, Appl. Opt. **16**, 24091 (1977)
35. A. C. Tam and C. K. N. Patel, Appl. Opt. **18**, 3348 (1979)
36. Z. Abdallah, Y. G. Boucher, A. Fernandez, S. Balac and O. Llopis, Sci. Rep. **6**, 27208 (2016)

37 D. Romanini, I. Ventrillard. G. Méjean, J. Morville, E. Kerstel, Cavity-Enhanced Spectroscopy and Sensing, ch. 1, G. Gagliardi & H. P. Loock eds., Springer-Verlag Berlin Heidelberg 2014
38 Y. He, B. J. Orr, Chem. Phys. Lett. **319**, 131 (2000)
39 M.-Y. Ye, M.-X. Shen, X.-M. Lin, Sci. Rep. **6**, 19597 (2016)
40 Y. Dumeige, S. Trebaol, L. Ghisa, T. K. Nguyen, H. Tavernier, P. Féron, J. Opt. Soc. Am. B **25**, 2073 (2008)
41 D. W. Vernooy, V. S. Ilchenko, H. Mabuchi, E. W. Streed, and H. J. Kimble, Opt. Lett. **23**, 247 (1998)
42 L. F. Phillips, J. Phys. Chem. B **108**, 1986 (2004)
43 D.G. Aarts, M. Schmidt, H.N.W. Lekkerkerker, Science **304**, 847 (2004)
44 J. Eggers, J. R. Lister, and H. A. Stone, J. Fluid Mech. **401**, 293 (1999)

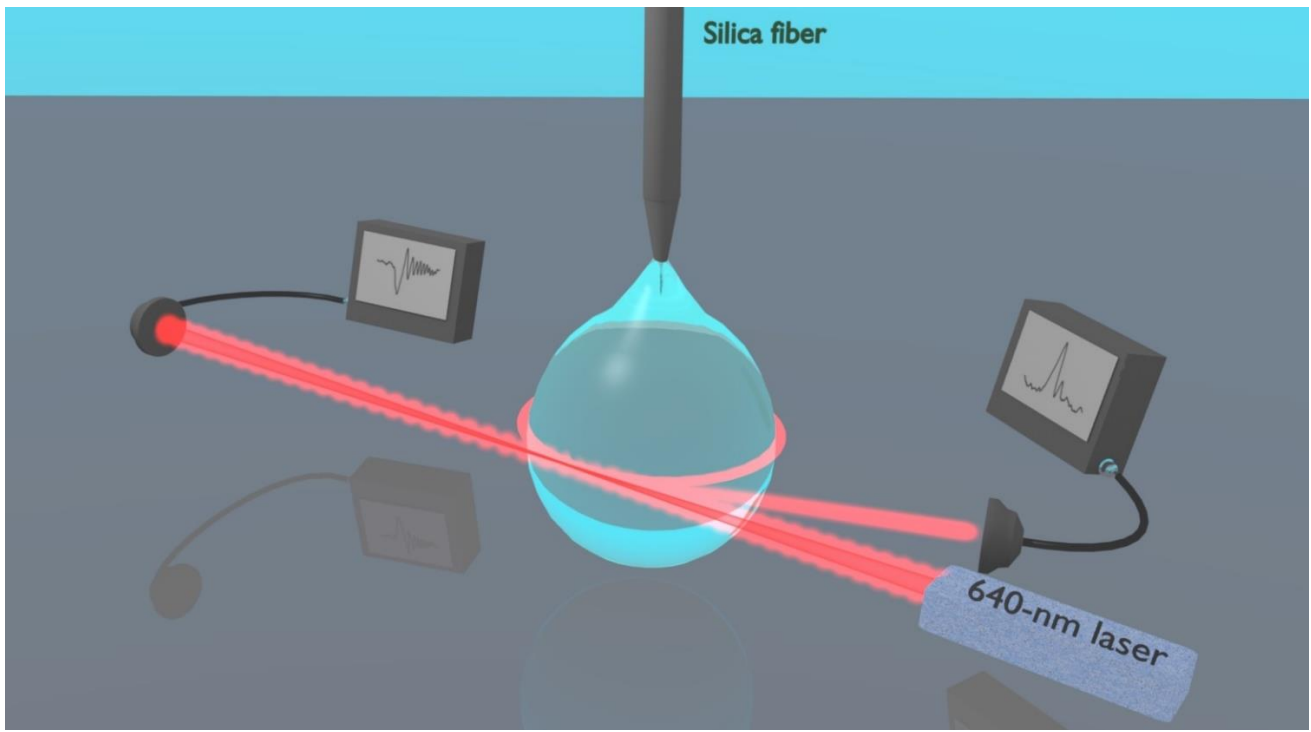

**Fig. 1**. Experimental arrangement. A visible diode laser is focused close to the droplet rim to excite WGMs. Ringing signals are observed where directly-transmitted light and the scattered light are detected.

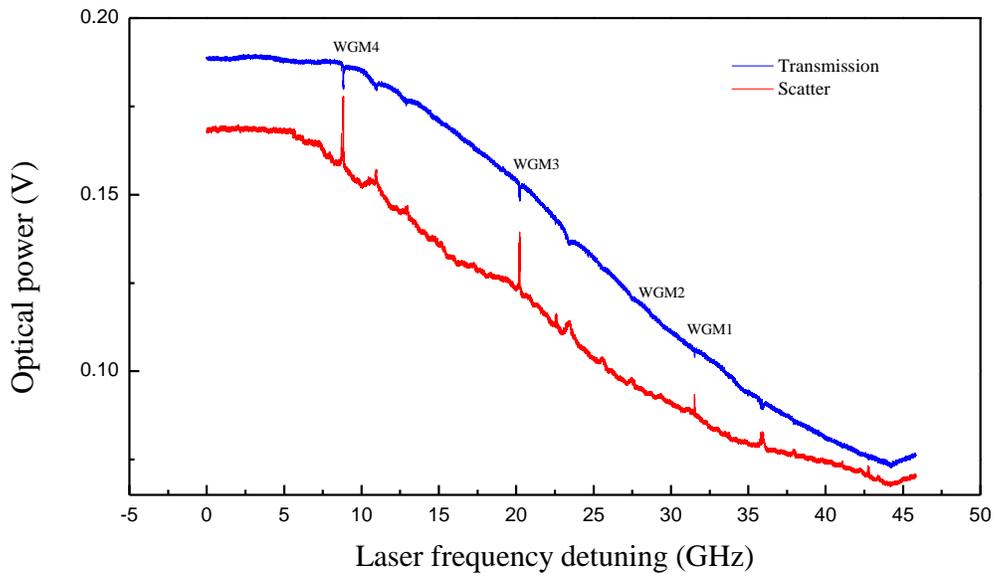

**Fig. 2**. WGM excitation along a laser-frequency sweep (13-Hz frequency) for a paraffin droplet (300-μm radius). The three radial modes excited by scattering show different linewidths. Lower-Q modes are visible on the background.

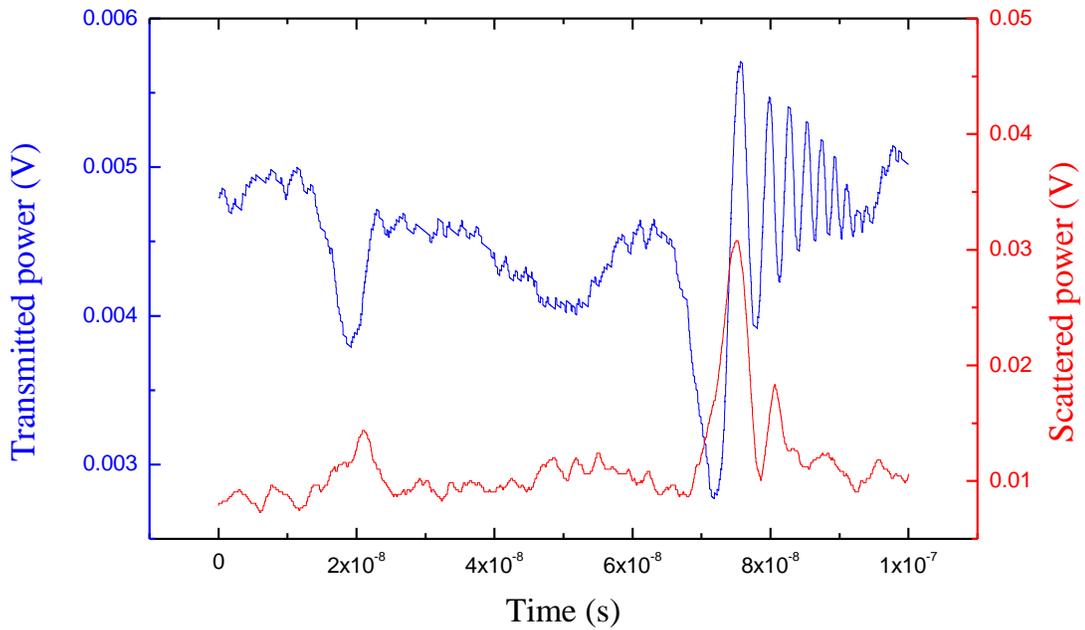

**Fig. 3.** Transient ringing waveforms recorded during a very-fast linear laser sweep (~ $10^{17}$ Hz/s) across the resonances corresponding to WGM1 and the weak mode WGM2.

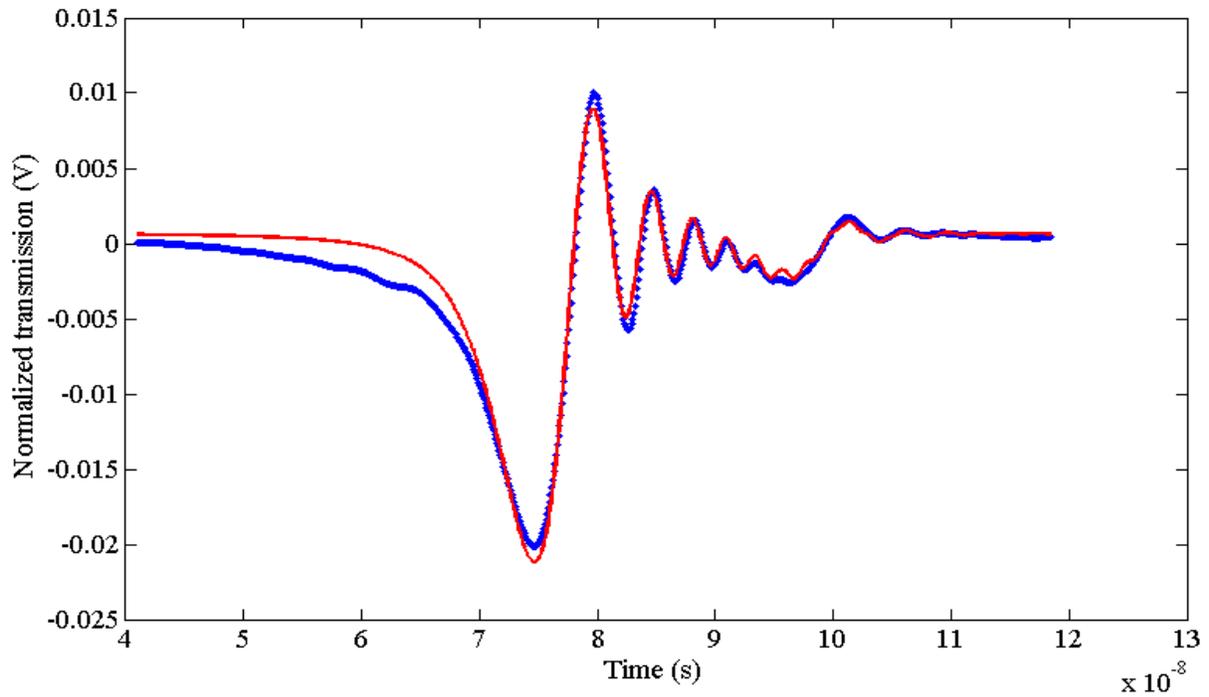

**Fig. 4**. Fitting of the WGM3 resonance with Eqs. 5 after normalization by the linear background. The overall lifetime is $\tau_3 = 4.8\pm0.1$ ns (errors are $1\sigma$ standard deviation). The time resolution is 5 ps.

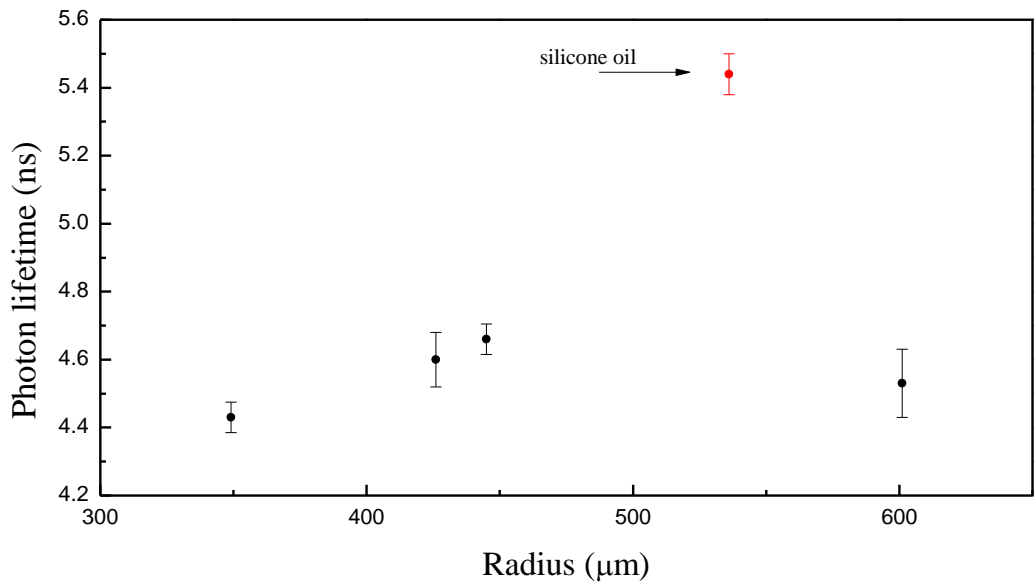

**Fig. 5.** Lifetime values extracted from the fit of ringing waveforms of paraffin droplets with different radii (the red point corresponds to a silicone oil droplet).

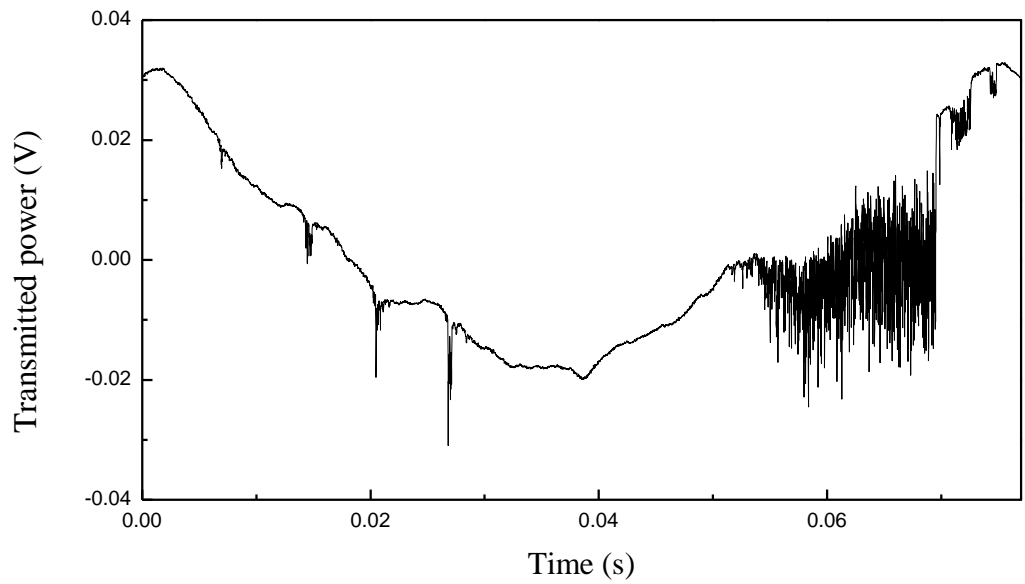

**Fig. 6.** Thermal broadening/narrowing effect and hysteretic response of a WGM resonance as seen on both sides of a slow linear wavelength scan (scan range ~ 1 GHz).